\documentclass[twoside,twocolumn,9pt]{article}
\usepackage{extsizes}
\usepackage[super,sort&compress,comma]{natbib} 
\usepackage[version=3]{mhchem}
\usepackage[left=1.5cm, right=1.5cm, top=1.785cm, bottom=2.0cm]{geometry}
\usepackage{balance}
\usepackage{mathptmx}
\usepackage{sectsty}
\usepackage{graphicx} 
\usepackage{lastpage}
\usepackage[format=plain,justification=justified,singlelinecheck=false,font={stretch=1.125,small,sf},labelfont=bf,labelsep=space]{caption}
\usepackage{float}
\usepackage{fancyhdr}
\usepackage{fnpos}
\usepackage[english]{babel}
\addto{\captionsenglish}{%
  
}
\usepackage{array}
\usepackage{droidsans}
\usepackage{charter}
\usepackage[T1]{fontenc}
\usepackage[usenames,dvipsnames]{xcolor}
\usepackage{setspace}
\usepackage[compact]{titlesec}
\usepackage{hyperref}

\definecolor{cream}{RGB}{222,217,201}

\begin{document}

\pagestyle{fancy}
\thispagestyle{plain}
\fancypagestyle{plain}{
%%%HEADER%%%
\renewcommand{\headrulewidth}{0pt}
}
%%%END OF HEADER%%%

%%%PAGE SETUP - Please do not change any commands within this section%%%
\makeFNbottom
\makeatletter
\renewcommand\LARGE{\@setfontsize\LARGE{15pt}{17}}
\renewcommand\Large{\@setfontsize\Large{12pt}{14}}
\renewcommand\large{\@setfontsize\large{10pt}{12}}
\renewcommand\footnotesize{\@setfontsize\footnotesize{7pt}{10}}
\makeatother

\renewcommand{\thefootnote}{\fnsymbol{footnote}}
\renewcommand\footnoterule{\vspace*{1pt}% 
\color{cream}\hrule width 3.5in height 0.4pt \color{black}\vspace*{5pt}} 
\setcounter{secnumdepth}{5}

\makeatletter 
\renewcommand\@biblabel[1]{#1}            
\renewcommand\@makefntext[1]% 
{\noindent\makebox[0pt][r]{\@thefnmark\,}#1}
\makeatother 
\renewcommand{\figurename}{\small{Fig.}~}
\sectionfont{\sffamily\Large}
\subsectionfont{\normalsize}
\subsubsectionfont{\bf}
\setstretch{1.125} %In particular, please do not alter this line.
\setlength{\skip\footins}{0.8cm}
\setlength{\footnotesep}{0.25cm}
\setlength{\jot}{10pt}
\titlespacing*{\section}{0pt}{4pt}{4pt}
\titlespacing*{\subsection}{0pt}{15pt}{1pt}
%%%END OF PAGE SETUP%%%

%%%FOOTER%%%
\fancyfoot{}
\fancyfoot[LO,RE]{\vspace{-7.1pt}\includegraphics[height=9pt]{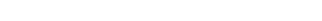}}
\fancyfoot[CO]{\vspace{-7.1pt}\hspace{13.2cm}\includegraphics{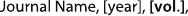}}
\fancyfoot[CE]{\vspace{-7.2pt}\hspace{-14.2cm}\includegraphics{head_foot/RF}}
\fancyfoot[RO]{\footnotesize{\sffamily{1--\pageref{LastPage} ~\textbar  \hspace{2pt}\thepage}}}
\fancyfoot[LE]{\footnotesize{\sffamily{\thepage~\textbar\hspace{3.45cm} 1--\pageref{LastPage}}}}
\fancyhead{}
\renewcommand{\headrulewidth}{0pt} 
\renewcommand{\footrulewidth}{0pt}
\setlength{\arrayrulewidth}{1pt}
\setlength{\columnsep}{6.5mm}
\setlength\bibsep{1pt}
%%%END OF FOOTER%%%

%%%FIGURE SETUP - please do not change any commands within this section%%%
\makeatletter 
\newlength{\figrulesep} 
\setlength{\figrulesep}{0.5\textfloatsep} 

\newcommand{\topfigrule}{\vspace*{-1pt}% 
\noindent{\color{cream}\rule[-\figrulesep]{\columnwidth}{1.5pt}} }

\newcommand{\botfigrule}{\vspace*{-2pt}% 
\noindent{\color{cream}\rule[\figrulesep]{\columnwidth}{1.5pt}} }

\newcommand{\dblfigrule}{\vspace*{-1pt}% 
\noindent{\color{cream}\rule[-\figrulesep]{\textwidth}{1.5pt}} }

\makeatother
%%%END OF FIGURE SETUP%%%

%%%TITLE, AUTHORS AND ABSTRACT%%%

%%%TITLE, AUTHORS AND ABSTRACT%%%
\twocolumn[
  \begin{@twocolumnfalse}
%{\includegraphics[height=30pt]{head_foot/SM}\hfill\raisebox{0pt}[0pt][0pt]{\includegraphics[height=55pt]{head_foot/RSC_LOGO_CMYK}}\\[1ex]
%\includegraphics[width=18.5cm]{head_foot/header_bar}}\par
\vspace{1em}
\sffamily
%\begin{tabular}{m{4.5cm} p{13.5cm} }

%\includegraphics{head_foot/DOI} &
 \noindent\LARGE{\textbf{Deformation profiles and microscopic dynamics of complex fluids during oscillatory shear experiments
}} %Article title goes here instead of the text "This is the title"
 \vspace{0.3cm} \\
 \noindent\large{Paolo Edera,\textit{$^{a}$} Matteo Brizioli,\textit{$^{a}$} Giuliano Zanchetta,\textit{$^{a}$} George Petekidis,\textit{$^{b}$} Fabio Giavazzi,\textit{$^{a}$} and Roberto Cerbino$^{\ast}$\textit{$^{a,c}$}} \\%Author names go here instead of "Full name", etc.

\noindent\normalsize{Oscillatory shear tests are widely used in rheology to characterize the linear and non-linear mechanical response of complex fluids, including the yielding transition. There is an increasing urge to acquire detailed knowledge of the deformation field that is effectively present across the sample during these tests; at the same time, there is mounting evidence that the macroscopic rheological response depends on the elusive microscopic behavior of the material constituents. Here we employ a strain-controlled shear-cell with transparent walls to visualize and quantify the dynamics of tracers embedded in various cyclically sheared complex fluids, ranging from almost-ideal elastic to yield stress fluids. For each sample, we use image correlation processing to measure the macroscopic deformation field, and echo-Differential Dynamic Microscopy to probe the microscopic irreversible sample dynamics in reciprocal space; finally, we devise a simple scheme to spatially map the rearrangements in the sheared sample, once again without tracking the tracers. For the yield stress sample, we obtain a wave-vector dependent characterization of shear-induced diffusion across the yielding transition, which is accompanied by a three-order-of-magnitude speed-up of the dynamics and by a transition from localized, intermittent rearrangements to a more spatially homogeneous and temporally uniform activity. Our tracking free approach is intrinsically multi-scale, can successfully discriminate between different types of dynamics, and can be automated to minimize user intervention. Applications are many, as it can be translated to other imaging modes, including fluorescence, and can be used with sub-resolution tracers and even without tracers, for samples that provide intrinsic optical contrast.} \\
 \end{@twocolumnfalse} \vspace{0.6cm}
]

%%%END OF TITLE, AUTHORS AND ABSTRACT%%%

%%%FONT SETUP - please do not change any commands within this section
\renewcommand*\rmdefault{bch}\normalfont\upshape
\rmfamily
\section*{}
\vspace{-1cm}

%%%FOOTNOTES%%%

\footnotetext{\textit{$^{a}$~Dipartimento di Biotecnologie Mediche e Medicina Traslazionale, Universit\`a degli Studi di Milano, via F.lli Cervi 93, 20090 Segrate, Italy}}
\footnotetext{\textit{$^{b}$~FORTH/IESL and Department of Materials Science and Technology, University of Crete, 71110 Heraklion, Greece}}
\footnotetext{\textit{$^{c}$~University of Vienna, Faculty of Physics, Boltzmanngasse 5, 1090 Vienna, Austria}}

%Please use \dag to cite the ESI in the main text of the article.
%If you article does not have ESI please remove the the \dag symbol from the title and the footnotetext below.
\footnotetext{\dag~Electronic Supplementary Information (ESI) available: [details of any supplementary information available should be included here]. See DOI: 10.1039/cXsm00000x/}
%additional addresses can be cited as above using the lower-case letters, c, d, e... If all authors are from the same address, no letter is required

%\footnotetext{\ddag~{Additional footnotes to the title and authors can be included \textit{e.g.}\ `Present address:' or `These authors contributed equally to this work' as above using the symbols:} \ddag, \textsection, and \P. {Please place the appropriate symbol next to the author's name and include a \texttt{\textbackslash footnotetext} entry in the the correct place in the list.}}

%%%END OF FOOTNOTES%%%

%%%MAIN TEXT%%%%
\section{Introduction}
As compared with molecular hard solids - such as steel, wood or bones - soft materials made of micron-sized constituents promptly deform with relatively small mechanical loads: we easily spread butter on a slice of bread, mindlessly distribute toothpaste on our teeth, effortlessly rub our faces with cream, smoothly tickle our palate with a chocolate mousse, and readily pull our cheeks with a little pinch. Beyond having practically ubiquitous consequences in our lives, this softness opens up unique possibilities when doing experimental research on soft matter. A paradigmatic example is \textit{passive microrheology}, in which the linear moduli of a soft material are probed by embedded microscopic tracer particles that operate as micro-scale rheometers powered by tiny thermal fluctuations \cite{waigh2016,edera2017}. Despite being a technique that has had enormous success, passive microrheology is unable to provide complete rheological information in all cases of interest. For instance, if the storage modulus of the material is too large or if one is interested in the non-linear rheological properties, it is often preferred to recur to \textit{active microrheology}, in which at least one of the embedded particles is driven by an external force obtained, for instance, \textit{via} optical or magnetic trapping \cite{waigh2016,robertson2018,valentine2012,vitali2021}. Passive and active microrheology techniques are very powerful tools for the characterization of soft materials, as they can access high frequencies and enable a space-resolved study of the rheological properties of complex, heterogeneous systems \cite{robertson2018}. 
However, the fact that the energy is injected at a local, particle scale makes them unsuitable to study and predict the complex failure modes observed in soft materials when they are mechanically perturbed at the macroscopic scale. 

More insightful information on the microscopic mechanisms underlying plasticity in soft materials can be obtained by imposing a controlled macroscopic deformation to the sample - for instance by displacing one or more of the confining surfaces of a shear rheometer - while monitoring over time the structural rearrangements occurring at the microscale. Rearrangements are typically monitored by accurate tracking of embedded tracer particles, performed in direct (rheo-microscopy or rheo-imaging) or in reciprocal space (rheo-scattering), very often but not necessarily by optical means \cite{aime2019,eberle2012,zanchetta2010exploring,saint2017}. These rearrangements may be partly or entirely reversible, with the irreversible component becoming dominant for increasing shear strain across the transition to non-linearity \cite{hebraud1997,petekidis2003,knowlton2014,rogers2014,aime2018}. This kind of experiments provides thus very useful insights on how the microscopic structure and dynamics determine the material rheology, which can be used to design enhanced materials, \textit{e.g.} with a longer lifetime, controlled relaxation, improved ductility or emerging memory \cite{nelson2017}.

A variety of shear rheology tests is compatible with the simultaneous characterization of the tracer particles' motion. The latter can be generally split into the sum of affine and non-affine displacements \cite{chen2010}; the relative importance of these two contributions depends on the type of test, and on the scheme used to sample the particles' configuration in space and time. For instance, during Small/Medium/Large Amplitude Oscillatory Shear (SAOS/MAOS/LAOS) rheology experiments, the so-called \textit{echo} protocol enables one sampling the microscopic configuration at each deformation cycle, by obtaining a stroboscopic movie of the particles configurations either in direct or in reciprocal space. In this way, the affine contribution is suppressed and one can focus on the non-affine irreversible particle rearrangements.

Most of the initial experiments with the echo protocol were performed with Diffusing Wave Spectroscopy (DWS)\cite{hebraud1997}, and Dynamic Light Scattering (DLS)\cite{petekidis2002}. Being based on the collection of multiply scattered light, DWS is very sensitive to a single, small length scale, whereas DLS is intrinsically a multi-scale technique, at the expense of an increased experimental complexity \cite{scheffold2007new}. More recently, microscopy has also been employed to track particle motion in oscillatory shear experiments with the echo protocol \cite{knowlton2014}.

If one is interested in going beyond the study of irreversible non-affine particle displacements typically obtained with the echo protocol, imaging and microscopy offer an important advantage over scattering experiments: the affine particle motion that characterizes the intra-cycle particle dynamics can be readily removed from the particle trajectories, whereas more caution is needed in scattering experiments \cite{aime2019II}. However, the information extracted from scattering experiments is characterized by an enviable statistical robustness of the obtained information. For these reasons, it would be interesting to have both techniques available during oscillatory shear experiments.

A possible compromise is represented by Digital Fourier Microscopy (DFM)\cite{giavazzi2014}, a family of techniques based on collecting in direct space temporal sequences of images of the sample, which are subsequently spatially Fourier transformed to obtain scattering information in the reciprocal space as a function of the wave-vector $\mathbf{q}$. The most popular DFM technique is Differential Dynamic Microscopy (DDM)\cite{cerbino2008,giavazzi2009,cerbino2017,cerbino2021differential}, which relies on subtracting images acquired at different times, a feature that appears well suited to echo experiments: when two images acquired at an integer multiple of the shear oscillation period are subtracted, only the irreversible particle displacements are captured and contribute to the final result. At the same time, one can take advantage of the availability of the direct space movies to remove the rigid translations associated to the affine intra-cycle particle dynamics\cite{cerbino2021}, a step that may benefit from the large amount of image registration algorithms available\cite{goshtasby2005}. %\textcolor{green}{Isn't this statement dangerous? Registration is effective in rigid drift subtraction, while, in general, the intra cycle dynamics are not affine, unless the depth-of-focus goes to zero or we are (at least approximately) in the echo condition.}

The potential of DDM in combination with shear rheology experiments was recently suggested in two companion articles by Aime and Cipelletti \cite{aime2019, aime2019II}, in which the authors performed a critical comparison with DLS for probing shear-induced rearrangements in Fourier space. In both cases, the intermediate scattering functions for $\mathbf{q}$ parallel to the direction of the applied shear (shear direction) are mostly dominated by the relaxation due to the affine particle displacements, whereas in the perpendicular direction (vorticity direction) only the non-affine particle rearrangements are probed. 
The main advantage of DDM over DLS is that the limited spatial coherence of illumination typical of bright-field DDM experiments results in a finite, $q$-dependent depth-of-focus, which can be used to probe slices of the sample perpendicular to the optical axis \cite{giavazzi2009}. This capability is surely appealing for those cases where the deformation field deviates from a simple homogeneous profile, such as in the presence of wall slip, shear banding, or edge fracture \cite{divoux2016}. All these different phenomena have the common effect of making the actual deformation field inhomogeneous and difficult to predict, causing problems in the interpretation of the results of both rheological and DLS experiments. Indeed, while in a rheological experiment, the characterization of the material constitutive equation for the shear stress $\sigma(\gamma)$, requires the accurate knowledge of the actual shear strain $\gamma$ of the sample, in DLS experiments, where the signal is the result of an average over the entire scattering volume, it is almost impossible to discern the contributions coming from different portions of a macroscopically inhomogeneous sample. Aime and Cipelletti have clearly and convincingly showed that a proper DDM characterization of the affine displacements is possible, providing that an instrument calibration of the $q$-dependent depth-of-focus is performed \cite{aime2019II}. However, the full potential of combining DDM analysis with an echo protocol, where the effect of affine displacements is minimized, remains yet unexplored. 

In this work, we perform oscillatory shear experiments with a strain-controlled cell, characterized by a parallel-sliding-plates geometry. During the sample oscillation, we monitor the dynamics of tracer particles by using bright-field DDM in conjunction with the echo protocol (echo-DDM). The depth-of-focus is tuned and optimized by varying the numerical aperture of the microscope condenser so that, even if the deformation field is not homogeneous across the whole gap, it is possible to selectively image narrow "slices" of the sample, where the shear can be assumed to be approximately constant (see Fig.\ref{fgr:setup}). We apply this optimization procedure to the imaging of various samples, some of which presenting heterogeneity of the deformation field in the direction of the velocity gradient (gradient direction). We perform a preliminary accurate characterization of the deformation field across the whole gap in standard samples with pure elastic solid or pure viscous liquid response, retrieving the expected linear deformation profile. By applying the same protocol to more complex soft matter samples (a viscoelastic fluid and a yield stress material), we demonstrate that our approach can also effectively probe deviations from such ideal profile, due for example to wall slip or shear banding. Quantification of these effects is key to correctly relate the microscopic irreversible dynamics to the local macroscopic deformation in echo experiments. 

Once obtained a reliable characterization of the local strain for all our samples, we use echo-DDM to probe the presence of a shear-induced non-affine rearrangements. The elastic solid, characterized by the absence of a true irreversible non-affine dynamics, serves as a reference to estimate the instrumental limits; as expected, Newtonian and Maxwell liquids do not exhibit any appreciable variation in non-affine activity as a function of the shear amplitude; by contrast, in the yield stress sample, we find intermittent, small scale displacements of the tracers for small strain amplitudes, and shear-induced diffusion \cite{miyazaki2004,evans1999,khabaz2021,mohan2013} for amplitudes larger than the one corresponding to the crossover point of the viscoelastic moduli. All the results obtained with echo-DDM are compared and found in agreement with particle tracking (PT) analysis of the tracer motion. Finally, we propose a simple yet robust scheme to draw activity maps highlighting the rearrangements that occur in the sheared sample.

The manuscript is organized as follows: in the Materials and Methods section we describe the tested samples (Sect \ref{samples}), the experimental set-up (Sect.\ref{expset}), and the experimental protocols enabling the characterization of the macroscopic deformation field (Sect.\ref{meso}) and of the irreversible non-affine dynamics (Sect.\ref{micro}). The results obtained in the characterization of the deformation profile and of the shear-induced rearrangements are presented and discussed in Sect.\ref{resdef} and Sect.\ref{resddm}, respectively, before presenting some concluding comments in Sect.\ref{conklu}.

%\textbf{DA QUI IN GIU' DA USARE ED ELIMINARE}
%\textcolor{violet}{Rheology of soft materials has both industrial and fundamental interest\cite{larson1999,sollich1997}}.
%\textcolor{violet}{Both for material design and for physical understanding one of the most interesting aspect is the interplay between the rheology and the microscopic structure and dynamics\cite{dreiss2007,mason1996}}.\\
%\textcolor{violet}{In the last decades the interest has moved to the dynamics\cite{petekidis2002, rogers2014,aime2018}}.\\
%\textcolor{violet}{To the best of our knowledge, previous works either focused on microscopic non-affine displacements, assuming a good control of the applied deformation field \cite{sentjabrskaja2015}, or focused on the deviation from affinity at a mesoscopic level, neglecting microscopic dynamics \cite{divoux2010,dinkgreve2018}}.\\

\section{Materials and Methods}
\label{MM}
\subsection{Samples}
\label{samples}
We test our experimental approach on samples that are representative of different classes of materials: an elastomer (Sylgard), a Newtonian fluid (silicon Oil), a Maxwell fluid (worm-like micelles, WLM), and a simple yield stress fluid (Carbopol). All the samples in our study are fairly transparent and are thus seeded with tracer particles of known size. However, seeding is not needed if the sample already provides a sufficient optical contrast, because DDM can capture the dynamics of a variety of samples including molecular mixtures \cite{buzzaccaro2013ghost,giavazzi2016equilibrium} and liquid crystals \cite{giavazzi2014viscoelasticity}.

Sylgard 184 (Dow Corning) is as a two-component system made of a silicon oil and a curing agent. A small fraction ($\phi \sim 0.1\%$) of sterically stabilized PMMA particles of diameter $260$ nm are dispersed in the silicon oil base before adding the curing agent. The sample is then mixed thoroughly, centrifuged to get rid of bubbles, and finally loaded in the shear-cell before the completion of the curing process, which takes about 48 h at room temperature. The so-obtained sample is expected to behave like a perfectly homogeneous elastic material over a wide range of shear strains\cite{aime2016}. The silicon oil that we use is the viscous component described above in the preparation of Sylgard. This sample behaves as a Newtonian viscous fluid with a shear viscosity $\eta \simeq 5$ Pa s \cite{SYLGARD}. Worm-like micelles are obtained by dispersing Hexadecylpyridinium chloride monohydrate (CpyCl) surfactant in water with Sodium Salicylate (NaSal) as a counter-ion\cite{dreiss2007,vitali2020}. Both components have been purchased from {Sigma-Aldrich} as powders. CpyCl is dispersed in Milliq water at a concentration 200 mM and taken to $40\ ^{\circ}C$  for 30 minutes, until it turns to a clear solution. NaSal is dispersed in Milliq water at a concentration of 120 mM. The two dispersions are slowly mixed in a 1:1 ratio and left on a rotating mixer overnight. The final molarities are 100 mM and 60 mM for CpyCl and NaSal, respectively. Polystyrene particles of 2 $\mu$m diameter (PS2, {Microparticles GmbH}) are dispersed at a volume fraction $\phi= 0.05\%$. The dispersion is initially mixed manually with a spatula and left on a rotating mixer for one night. Carbopol 971 P NF has been purchased from Lubrizol as a single component powder. The powder is dispersed in MilliQ water to obtain a concentrated starting solution with concentration $c=5\%$ wt. The obtained dispersion has an acidic pH, that is neutralized by adding few drops of NaOH 10 M to 10 ml of concentrated dispersion, while gently stirring \textcolor{black}{for several days by keeping the pH controlled. At the end of the mixing phase,} the neutralized dispersion is then diluted to obtain the sample at the desired concentration $c=0.5\%$ wt. \textcolor{black}{We found that this preparation procedure ensures the reproducibility of the rheological properties of the Carbopol samples}. Eventually, PS2 tracer particles are dispersed at a volume concentration $\phi= 0.05\%$. 
Before the experiment, all samples are centrifuged to get rid of the air bubbles. \textcolor{black}{We note that the choice of the tracer particle size for the Carbopol sample was guided by preparatory experiments in which we investigated the tracer dynamics in Carbopol at rest for different particle diameters: $0.5$, $0.8$, $2$, and $5$ microns. For the $0.5$ and $0.8$ micron particles, we found that roughly half of the particles appeared stuck and the other half underwent caged dynamics. By contrast, for the $2$ and $5$ micron tracers, all the particles were substantially trapped. These observations, in agreement with seminal microrheological studies of Carbopol EDT 2050 \cite{oppong2006microrheology}, point to the existence of a characteristic microstructural length scale, which in our Carpobol sample is smaller than 2 micron.}
\subsection{Experimental Setup}
\label{expset}
\subsubsection{Shear-cell}
\label{shear-cell}
The samples are loaded between the two horizontal parallel plates of a commercial strain-controlled shear-cell \textcolor{black}{(RheOptiCAD, CAD Instruments, France)}.
With this shear-cell, {a modified version of the one} described in Ref. \cite{boitte2013} to enable bright-field imaging, 
we can impose an oscillatory displacement of the top plate
\begin{equation}
\Delta \mathbf{X}_0(t) =\text{ } A_0 \sin{(2\pi \textrm{f} t)} \hat{u}_x,
\label{kapparoger}
\end{equation}
while the bottom plate remains stationary. In our experiments, the oscillation frequency is kept constant ($\textrm{f}=1$ Hz), while the amplitude $A_0$ is varied in the range $[0.01-1]$ mm. The displacement of the top plate identifies the shear direction, corresponding to the orientation of the $x$-axis in Fig.\ref{fgr:setup}. The height of  gap between the plates is 1  mm for Sylgard, 250 $\mu$m for Worm-like micelles, and  500 $\mu$m for the other samples. The plates are standard glass slides whose surface was sandblasted to reduce wall slip, except for small a circular window of diameter 4 mm, which is left clear to enable optical access to the sample. \textcolor{black}{Even though we have not conducted an independent characterization of the roughness of the sandblasted surfaces, we expect it to be of a few micrometers, slightly larger than both the microstructure length-scale of the Carbopol and the particle size.} The orientation of the plates is controlled by micrometer screws. The parallelism of the two plates and the alignment of the moving plate along the shear direction are essential to produce a simple shear deformation.
The proper alignment of the top plate is assessed by observing the plate surface under the microscope while it is displaced along the shear direction and checking whether it remains in sharp focus.
%To this aim, we image with the microscope on the top plate and we check that moving it to its extremes in the shear direction doesn't produce a displacement in the $z$ direction (taking it out of focus).
As described in more detail in {ESI}, the angle between the two plates is measured by means of a custom interferometric system, coupled to the rear port of the microscope. The alignment accuracy than can be achieved is within $\delta \theta \sim 10^{-3}\ rad$.

Due to a memory buffer register limitation, with the current version of the shear-cell control software the length of an oscillatory experiment is limited to $500$ cycles. %,  this limits the duration of the experiments on Sylgard, where mechanical instability are negligible. As discussed in the following, in fluid-like samples the duration of the Echo DDM analysis is constrained by the presence of drifts.\\
Moreover, the shear-cell does not provide a real-time output signal proportional to the position of the moving plate, which could have been used to trigger a synchronized camera acquisition. Synchronization is a crucial issue in our set-up and it  will be discussed in detail in Sec. \ref{technical}.

\subsubsection{Microscope}
The shear-cell described in the previous section is mounted on an inverted microscope ({Nikon Eclipse Ti-S}), equipped with a 20x 0.45 NA objective. The working distance of the objective (WD= 8 mm) is large enough to enable scanning the whole gap between the two plates confining the sample. As discussed also in the the ESI, the vertical resolution in the reconstructed deformation field is determined by the depth-of-focus of the imaging system. In our setup, we minimize the depth-of-focus (to about 20 $\mu$m, see {ESI}  for details) by keeping the condenser diaphragm  almost completely open (LWD, NA=0.52).

%\subsection{Experimental procedure and data analysis}
\subsection{Deformation field}
\label{meso}
%\textcolor{green}{The first point that we address concerns the possible deviations of the deformation field from homogeneity.}

\begin{figure}[h]
\centering
  \includegraphics[width=9.5 cm]{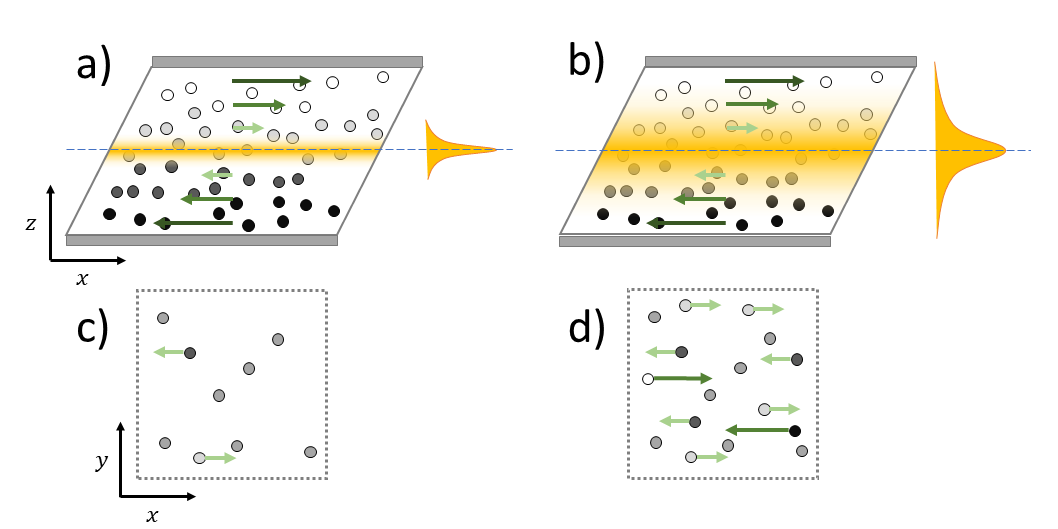}
  \caption{\textbf{Effect of the depth-of-focus in optical microscopy.} (a) Schematic representation of a sheared sample in sliding parallel plates geometry, in the shear-gradient ($x$-$z$) plane.  Green arrows represent local displacements with respect to the rest position, tracer particles are color-coded (from black to white) according to their vertical position. The dotted line corresponds to the intersection with the object plane of the imaging system, whose optical axis coincides with the velocity gradient ($z$) direction. The yellow shaded region, whose width coincides with the depth-of-focus $L_f$, identifies the portion of the sample contributing to the image (c).
  Panel (b) depicts the same sheared sample shown in panel (a), observed with a larger depth-of-focus. The corresponding image, shown in panel (d), features a larger number of particles and is characterized by a larger dispersion in the particle displacements along the shear direction. }
\label{fgr:setup}
\end{figure}

It is known that in several conditions of interest, the deformation field within a sheared sample may be non-homogeneous even when a uniform stress is applied \cite{dinkgreve2018}. Two examples of deviation from homogeneity are shear banding and wall slip. These phenomena affect a wide variety of samples \cite{vasisht2020} \cite{barlow2020} \cite{divoux2010} and while they have an indisputable effect on the mechanical response and thus on the interpretation of rheological data, it is not easy to pinpoint their presence from purely mechanical measurements. 
In the following, we provide a simplified description of these effects in the case of the parallel sliding plates geometry, which is the one used in our experiments and is schematically represented in Fig.\ref{fgr:setup}. A key advantage of this geometry, compared to other common configurations like rotating plate-plate or rotating cone-plate, is that it allows imposing a uniform strain at the confining surfaces while ensuring easy optical access to the sample \cite{aime2019,rogers2014,knowlton2014}.

Ideally, when a uniform displacement $\Delta \mathbf{X}_0=A_0 \hat{u}_x$ is imposed on the top plate, while the bottom one is fixed, an affine displacement field is produced in the sample
\begin{equation}
    \Delta \mathbf{X} (z) = z\ \gamma_0 \  \hat{u}_x,
\label{DF}
\end{equation}
where $z$ is the vertical coordinate (corresponding to the velocity gradient direction), and has the origin of the axis at the bottom plate; $\hat{u}_x$ in the unit vector oriented along the $x$ (shear) direction and the ratio $\gamma_0= A_0/h$ between the imposed displacement at the moving plate and the gap width $h$ is the imposed shear strain. The displacement is uniform over each horizontal plane, with no components along other directions.
The corresponding  strain, which is given by the spatial gradient of the displacement field ${\nabla}\big( \Delta \mathbf{X}\big)$, is perfectly homogeneous and the only non-zero component of the strain tensor corresponds to the shear strain 
\begin{equation}
   \mathbf{\gamma} = \frac{\partial{(\Delta \mathbf{X} \cdot \hat{u}_x})}{\partial z}  =  \gamma_0\ .
    \label{dispdef}
\end{equation}

Deviations from this ideal case occur for example in case of wall slip, when the {adhesion} between the sample and the confining plates is not sufficient to prevent their relative motion. When wall slip occurs, in the absence of other effects, like shear-banding, Eq.\ref{DF} modifies to
\begin{equation}
    \Delta \mathbf{X} (z) = (z - z_0)\ \gamma_0 c\  \hat{u}_x,
\label{DFS}
\end{equation}
where $z_0$ is the coordinate of the horizontal plane (outside the sample) on which the extrapolated displacement profile would vanish, and $c<1$. In this scenario, the strain field is still homogeneous, even if with the actual shear strain $\gamma$ is smaller than the imposed one $\gamma=c\gamma_0<\gamma_0$. Since the actual deformation is obtained multiplying the imposed one by $c$, we define $c$ as the {deformation-loss-factor}. \textcolor{black}{Eq. \ref{DFS} can reproduce any (symmetric or asymmetric) linear profile with “perfect slip” at one or two surfaces, by suitably choosing the parameters $0\le c\le 1 $ and $-h\le z_{0}\le 0$.}\\
Another important example of deviation from homogeneity of the deformation field is shear banding, corresponding to the separation of the sample into macroscopic regions (bands) with different local shear strains \cite{divoux2010, ovarlez2009}. As an example, one can consider the simplistic case of a system separated in two bands, in the absence of wall slip, which is described by the displacement profile
\begin{equation}
    \Delta \mathbf{X} (z)  =
    \begin{cases}
                   \ z\ \gamma_1\  \hat{u}_x    	& 	\text{for   } z\leq z_1\\
                  \Big[\ (z-z_1)\ \gamma_2\ + z_1\ \gamma_1 \Big] \hat{u}_x  	&  	\text{for   } z>z_1
               \end{cases},
\label{DDDFS}
\end{equation}
where $\gamma_1$ and $\gamma_2$ are the local shear strains in the two bands. Even in this simple case where just two bands are formed, many profiles are compatible with a given imposed deformation $\gamma_0$, corresponding to all possible choices of the parameters $z_1$ and $\gamma_1$ satisfying the constraint $z_1\gamma_1\leq h\gamma_0$. Under these conditions, even if the average shear strain is bound to be $\gamma_0$, the local deformation can be very different, assuming in principle any value from $0$ to infinity, which reinforces the importance of mapping the local strain. \textcolor{black}{Eqs. \ref{DFS} and \ref{DDDFS} can be easily combined to obtain a description of the deformation field with coexisting wall slip and shear banding. The simple model described by Eq. \ref{DDDFS} involves a sharp discontinuity of the local strain that may be avoided with more complex modelling in which the local strain changes continuously \cite{seth2012soft,divoux2015wall}. For all the experiments described in this work, Eq. \ref{DFS} provides a satisfactory description.
}

In our bright-field microscopy experiments, we exploit the optical sectioning capability resulting from the partial coherence of illumination to obtain an accurate characterization of the displacement $ \Delta \mathbf{X} (z) $ at different heights $z$, and thus a faithful reconstruction of the actual shear strain profile $\mathbf{\gamma}(z)$ within the sample. 

\subsubsection{Strain field under oscillatory shear}
\label{strainfieldmeas}
% To maximise the accuracy of the measurement we want to minimize the depth-of-focus $L_f$. One image is in fact a projection with  depth $L_f$ around the focal plane, and the depth-of-focus $L_f$ can be tuned operating on the numerical apertures of the illumination source and on the objective.
We impose an oscillatory displacement of the top plate described by Eq.\ref{kapparoger}, with imposed amplitude $A_0$ and shear strain $ \gamma_0 = A_0/h$. We then perform observations of the sheared sample at different vertical positions $z_i \in [0,h]$. For each height $z_i$, a stack of 200 images is acquired with a frame rate $1/\delta t = 100$ fps (see {ESI}).
%\textcolor{black}{PAOLO: are we sampling just 2 complete oscillations?}\textcolor{green}{Sì. Sampling only two oscillations per height we limit the duration of a complete vertical scan. The vertical scan is normally repeated two times to exclude that the strain field evolves in time.}

%, number of images $200$.

\begin{figure}[h]
\centering
  \includegraphics[width=9 cm]{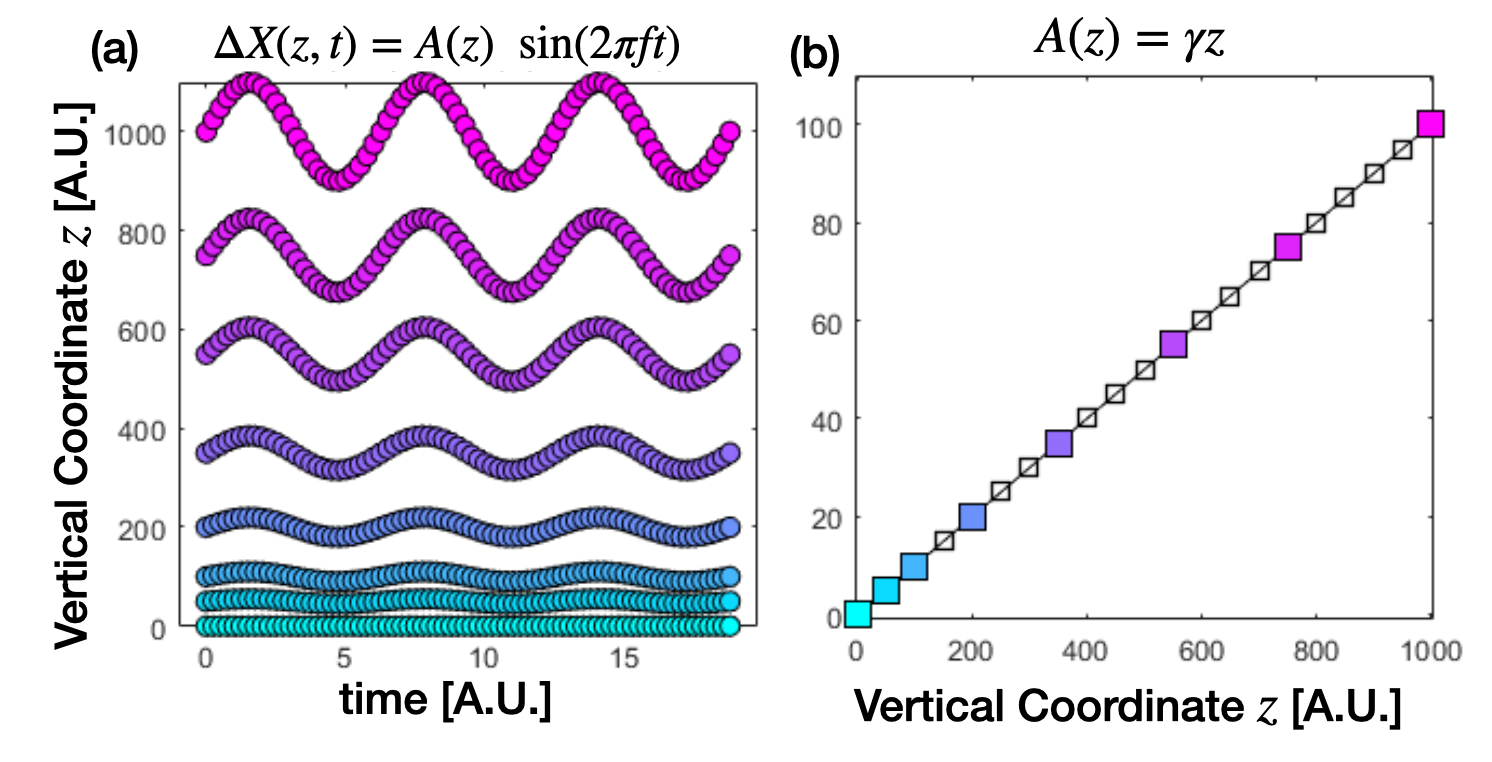}
  \caption{\textbf{Determination of the local deformation field.}  (a) Displacement field $\Delta \mathbf{X}(z_i,t) = A(z_i) \sin{(2\pi \textrm{f} t)} \hat{u}_x$ at different vertical coordinates $z_i$ in a simulated oscillatory strain experiment in the sliding parallel plates geometry in the absence of wall slip and shear banding. From a sinusoidal fit of the displacement field at each  given height $z_i$ one can obtain the corresponding amplitude $A(z_i)$ (b). The local shear strain can be then estimated as $\gamma(z)= \partial_z A(z)$, which becomes z-independent for the example considered in a).}
  \label{fgr:def}
\end{figure}

We calculate  the spatial cross-correlation function $c(\mathbf{r},t)$ of each pair of consecutive frames collected at a given height %$z_i$
\begin{equation}
c(\mathbf{r},t)= \Sigma_{\mathbf{x}} I( \mathbf{x},t+\delta t)\cdot I(\mathbf{x}+ \mathbf{r},t).
\end{equation}
The position $ \mathbf{\bar{r}}(t)$ of the peak of $c(\mathbf{r},t)$ provides an estimate for the rigid displacement occurred between the two images, taken at times $t$ and $t+\delta t$, respectively. By summing up all contributions from consecutive image pairs, we reconstruct the full displacement profile $\Delta \mathbf{X}(z_i,t) =\sum_{t^\prime\leq t} \mathbf{\bar{r}}(t^\prime)$ at the considered vertical position, as schematically shown in Fig.\ref{fgr:def}(a). Fitting of this profile to a sinusoidal function enables extracting the corresponding amplitude $A(z_i)$ (see Fig.\ref{fgr:def}(b)). We then estimate the local shear strain by numerically evaluating the partial derivative of the displacement field along the vertical direction $\gamma(z)=\partial{A_0(z)}/\partial z $. 
During an oscillatory shear experiment, we perform repeated scans at different times to monitor a possible time evolution of the deformation profile (further detains in the ESI).

\subsection{Microscopic non-affine displacements}
\label{micro}
The emergence of both reversible and irreversible microscopic non-affine displacements has been reported in simulations and experiments on sheared soft solids, and it has been related to the transition from linear to non-linear mechanical response, as well as to their yielding or flow behavior \cite{goldenberg2007,petekidis2002, mohan2013, colombo2014stress}.  %they are expected to play a central role as the shear perturbation is such to take the system out of its linear response region. % 
In such cases, even if the macroscopic displacement field $\Delta \mathbf{X}$ remains homogeneous (see Eq.\ref{DF}), the actual displacement field:
\begin{equation}
    \Delta \mathbf{x}(x,y,{z},t) =   \Delta \mathbf{X}({z},t)+\Delta \mathbf{x}_{N.A.}(x,y,{z},t).
\label{NAD}
\end{equation}
includes also a non-affine contribution $\Delta \mathbf{x}_{N.A.}$ due to rearrangements occurring on the scale of the elementary building blocks of the material.
In general, $\Delta \mathbf{x}_{N.A.}$ is  inhomogeneous in the shear-vorticity plane, and it is not necessarily oriented along the shear direction. In the rest of Sect.2, after briefly reviewing the working principle of DDM, we discuss how it can be adapted to capture and quantify irreversible non-affine displacements in a periodically sheared sample. 

\subsubsection{Differential Dynamic Microscopy}
\label{DDM}
\begin{figure}[h]
\centering
  \includegraphics[width=8 cm]{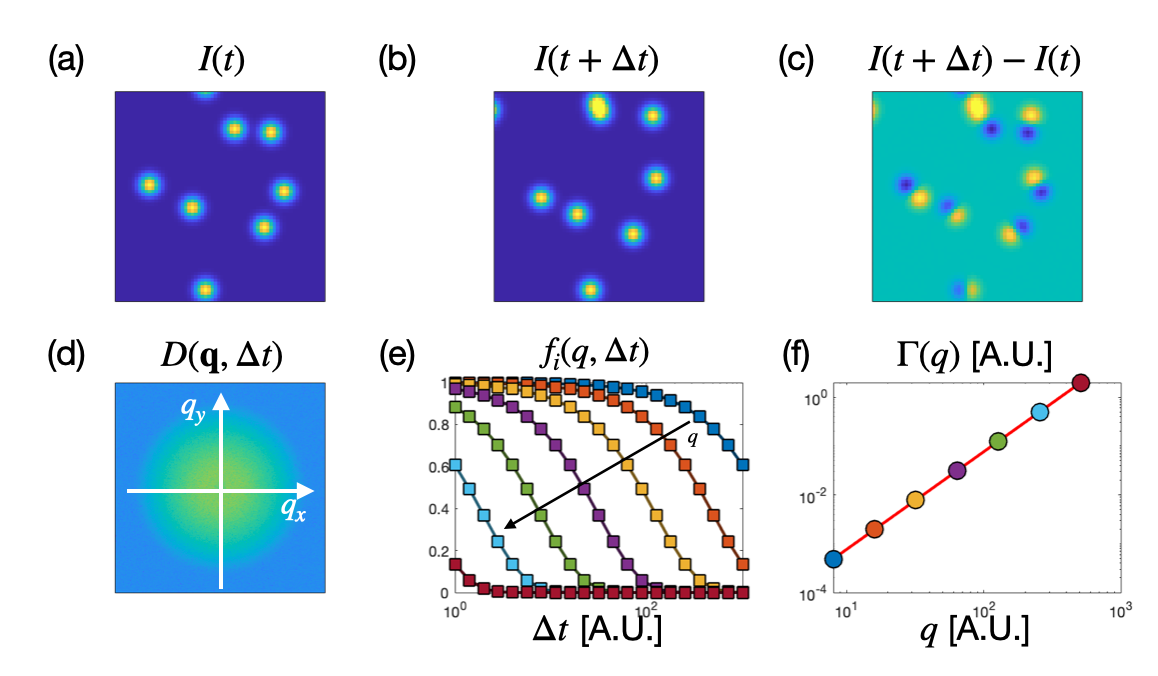}
  \caption{\textcolor{black}{\textbf{Differential Dynamic Microscopy workflow}. Synthetic images of particles (each particle has a Gaussian intensity profile) corresponding to time $t$ (a) and $t+\Delta t$ (b), and their difference (c). Pictorial representation of the two-dimensional dynamic structure function $D(\mathbf{q},\Delta t)$ (d), and of the azimuthally-averaged intermediate scattering functions $f_{i}(q,\Delta t)$ obtained at different wave vectors q (e). Fitting of the ISFs with a suitable model (continuous curves in panel e) enables extracting a $q$dependent relaxation rate $\Gamma(q)$ (f).}}
  \label{fgr:DDMex}
\end{figure}

The first step of the DDM analysis is the calculation of the difference between two images acquired at times $t$ and $t+\Delta t$, respectively (see Fig.\ref{fgr:DDMex}a-c). As proposed in Ref. \cite{giavazzi2017}, we then multiply the so-obtained image difference with a windowing function before calculating its Fourier power spectrum and successively average over all the results obtained with the same time delay $\Delta t$ but different initial times $t$. The result of this average is the image structure function (Fig.\ref{fgr:DDMex}d)
\begin{equation}
    D(\mathbf{q}, \Delta t)= \langle|\hat{I}(\mathbf{q}, t+\Delta t) - \hat{I}(\mathbf{q}, t)|^2\rangle,
    \label{powerspec}
\end{equation}
where $\hat{I}(\mathbf{q}, t)= \text{FFT}\{ I(\mathbf{x}, t) \}$ is the Fast Fourier Transform of the image at time $t$, $\mathbf{q}$ is the wavevector, and the symbol $\langle \cdot \rangle$ indicates an average over the initial time $t$. This calculation is iterated for all the time delays $\Delta t$ that are needed to properly follow the sample relaxation. 
The interpretation of the results is based on the fact that $D(\mathbf{q},\Delta t)$ can be written in terms of the real part  $f(\mathbf{q},\Delta t)$ of the intermediate scattering function (ISF), which is the quantity encoding information on the relaxation of the density Fourier modes, and whose squared amplitude is typically measured with DLS experiments:   \cite{cerbino2008}
\begin{equation}
    D(\mathbf{q},\Delta t)= a(\mathbf{q}) [ 1-f(\mathbf{q},\Delta t)]+b(\mathbf{q}).
    \label{dynamicf}
\end{equation}
The amplitude term $a(\mathbf{q})$ is determined by the scattering properties of the sample and by the microscope transfer function, whereas the additive term $b(\mathbf{q})$ accounts for the noise in the detection chain \cite{giavazzi2014}. To extract the ISF from the image structure function, $a(\mathbf{q})$ and $b(\mathbf{q})$ need to be determined.
Since $f(\mathbf{q},\Delta t\rightarrow0)=1$, $b(\mathbf{q})$ could be in principle estimated as the limit for small $\Delta t$ of $D(\mathbf{q},\Delta t)$. In practice, we fit a quadratic function to the image structure function over a small-time interval close to the origin (typically including the first five delay times) and we estimate $b(\mathbf{q})$ as the intercept of the best fitting curve \cite{Cerbino_dark_2017}.
Similarly, since $f(\mathbf{q},\Delta t\rightarrow \infty)=0$,  one could think to extract $a(\mathbf{q})$ from the long-time plateau of  $D(\mathbf{q},\Delta t)$. Unfortunately, because of the finite acquisition time set by the experimental constraints, in many cases we are not able to capture the full relaxation of $f(\mathbf{q},\Delta t)$ for all wavevectors. Consequently, we decided to adopt a different strategy \cite{Cerbino_dark_2017}. We identify a background image $I_0(\mathbf{x})$ incorporating all the static, sample-independent contributions to the images (see {ESI} for detail) and we estimate the amplitude $a(\mathbf{q})$ directly from the power spectrum of the raw images, corrected for the background  $a(\mathbf{q}) = 2\langle |\hat{I}(\mathbf{q},t)-\hat{I}_0(\mathbf{q})|^2 \rangle-b(\mathbf{q})$.

Once $a(\mathbf{q})$ and $b(\mathbf{q})$ are known, Eq.\ref{dynamicf} can be inverted to obtain the ISF. The sample dynamics can be then extracted, for example, by assuming a specific model for the statistics of the particle displacements and fitting it to the ISF. For a dispersion of identical, non-interacting Brownian particles in a Newtonian fluid, the ISF is azimuthally symmetric and is given by $f(q,\Delta t)=e^{-D_0q^2\Delta t}$, where $D_0$ is the diffusion coefficient of the particles. In this case, $D_0$ can be estimated as a coefficient of the $q$-dependent relaxation rate $\Gamma(q)=D_0q^2$, obtained by fitting an exponential function $e^{-\Gamma(q)\Delta t}$ to the ISF for different wavevectors (see Fig.\ref{fgr:DDMex}(e-f)). 
More generally, at least in the case of identical, non-interacting particles, the IFS can be expressed in terms of the Fourier transform of the probability distribution function of particle displacements $P(\mathbf{r},\Delta t)$, \textit{i.e.} the self part of the van Hove function \cite{hansen2013theory}
\begin{equation}
    f(\mathbf{q},\Delta t)= \int d^2 \mathbf{r} P( \mathbf{r},\Delta t) e^{-j\mathbf{q}\cdot\mathbf{r}}.
    \label{ISF_FT}
\end{equation}
If the dynamics is anisotropic, it can be worth considering separately the behavior of the ISF along specific directions in Fourier space.
In this case, the azimuthal average of $f(\mathbf{q},\Delta t)$ in the $\mathbf{q}$-plane can be restricted to narrow angular sectors of aperture $\Delta \theta$ oriented along the direction of interest\cite{giavazzi2014viscoelasticity}.
Under the same conditions under which Eq.\ref{ISF_FT} holds, we can write
\begin{equation}
    \langle|\Delta x(\Delta t)|^2\rangle=-\frac{\partial^2}{\partial q_x^2} f(\mathbf{q},\Delta t)|_{\mathbf{q}=0},
    \label{ISF_MSD}
\end{equation}
which enables estimating the mean square displacement along a selected direction (say, the $x$-axis) directly from the low-$q$ behavior of the ISF along the corresponding direction in the Fourier space.

\subsubsection{Echo-Differential Dynamic Microscopy (Echo-DDM)}
\label{EDDM}
In the presence of an imposed oscillatory deformation, an effective strategy to single out shear-induced irreversible rearrangements is to adopt a stroboscopic or echo approach, where the microscopic configuration of the system is probed at a sampling frequency that matches the oscillation frequency. This approach to shear-induced dynamics has been previously exploited in combination with light scattering, microscopy \textcolor{black}{or ultrasonic echography} \cite{hebraud1997,PhysRevLett.79.1154, petekidis2002, gibaud2010heterogeneous,knowlton2014}. In the case of microscopy, one considers sequences of echo images $I_n(\mathbf{x})$ taken at times $t_n=t_0+nT$, where $t_0$ is a time offset, $T$ is the oscillation period and $n$ is an integer number comprised between 1 and the total number of cycles $N$.
This choice ensures that each frame is captured in correspondence to the same realization of the macroscopic displacement field and, as a consequence, any difference between the images can be attributed to the occurrence of non-affine, irreversible displacements within the sample.
 Since DDM is a very sensitive probe of the microscopic dynamics, one could think to use it directly to extract the effective, shear-induced dynamics present in the echo images.
However, testing this idea requires first minimizing a number of potential artifacts which, if not adequately accounted for, can introduce spurious contributions to the echo dynamics.
These artifacts can arise for example from drifts or flows due to an imperfect confinement of the sample or from a mismatch between the oscillation frequency and the sampling frequency.
In the following section, we present a pre-processing procedure aimed at minimizing such effects and enabling robust DDM analysis.

\subsubsection{Correction of real and apparent drifts}
\label{technical}
\begin{figure}[h]
\centering
  \includegraphics[width=8 cm]{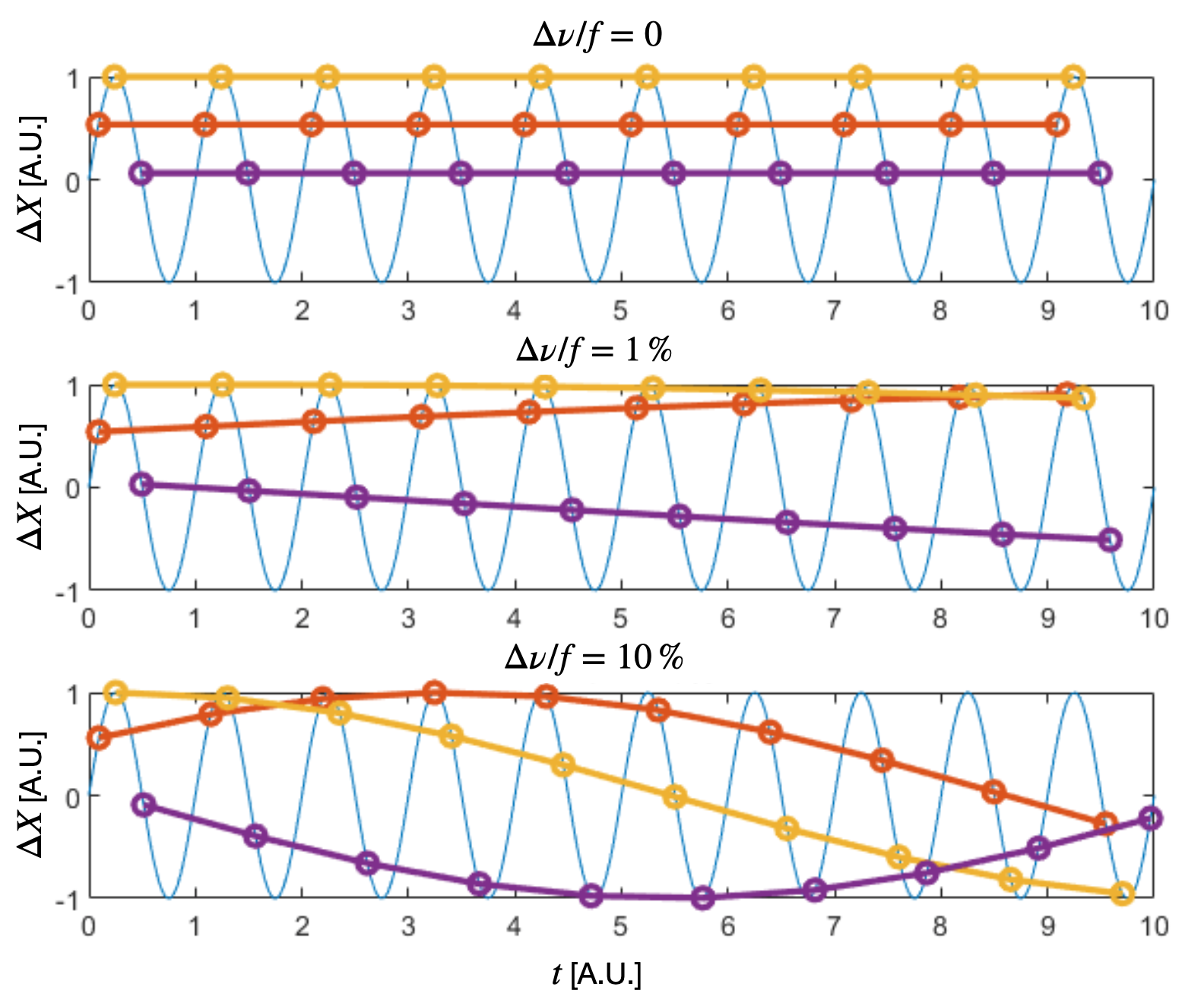}
  \caption{\textbf{Apparent displacement due to imperfect synchronization between deformation and acquisition}. In each panel, the continuous blue line represents the displacement along the shear direction of the observed plane as a function of time $\Delta X(t)=\sin{(2\pi \textrm{f} t)}$, the circles indicate the position at the sampling times $t_n=(\varphi_i/2\pi + n )/\nu$, where $\nu=\textrm{f}+\Delta \nu $ is the sampling frequency and $n$ is a positive integer number. Circles of different colors correspond to different initial phases $\varphi_i$.
  In each panel, a different value of the frequency mismatch $\Delta \nu$ is considered $\Delta \nu/\textrm{f}=0, 1\%$, and  $10\%$, respectively. 
If $\Delta \nu = 0$ each sampling series always captures the system in the same position. When $\Delta \nu \neq 0$ instead, an apparent displacement is observed in each sampling series, the velocity of the apparent displacement depending on both $\Delta \nu$ and on the initial phase $\varphi_i$.}
  \label{fgr:beat}
\end{figure}

As observed in previous work \cite{knowlton2014}, the imperfect synchronization between deformation and acquisition can lead to an apparent, phase-dependent displacement in the echo images, possibly impairing the result of the experiment (see Fig.\ref{fgr:beat}).
A relatively simple solution for obtaining a good synchronization would be to trigger the camera acquisition with a signal obtained from the shear-cell, proportional to the displacement imposed on the upper plate. Unfortunately, our shear-cell does not provide such a real-time output signal \cite{boitte2013}, which forced us to act both at the instrumental level and at the image processing level.
The proposed solution encompasses the following steps: accurate characterization of the oscillation frequency of the shear-cell, adoption of the best-matching image acquisition frequency, measurement of the residual drift in the collected images, and correction of the residual drift \textit{via} image registration. We now describe these corrective steps.

\textbf{Characterization of the oscillation frequency}. In our set-up, the shear-cell and the camera are controlled by two different computers. The first step is to characterize the frequency of the shear-cell in terms of the clock of the camera's computer. We observed that the actual frequency of oscillation is systematically different from the nominal one in an amplitude-dependent fashion, so that the actual frequency $\textrm{f}$ turns out to be a function of the two input parameters $\textrm{f}(\textrm{f}_0,A_0)$. This function for $\textrm{f}_0=1\ Hz$, $A_0\ \in (10\ \mu m, 1\ mm)$ is reported in ESI (Fig.S2).  The relative difference between the nominal and actual oscillation frequency  $(\textrm{f}-\textrm{f}_0)/\textrm{f}_0$ is typically or the order of $0.1\%$. This difference turns out to be significant, especially for long experiments lasting some hundreds of oscillations.
The experiment-to-experiment variability is estimated with the root mean square deviation measured for fixed input parameters $(\textrm{f}_0,A_0)$, $\bar{ \delta} \textrm{f} \sim 1 \cdot 10^{-5} \text{Hz}$ (details of the frequency characterization are given in ESI).

\textbf{Frequency matching:}
To increase the statistics while keeping  the duration of the experiments fixed, it is convenient to acquire more than one image per period. Thus the acquisition frequency $\nu$  has to meet the condition  $\nu =m_{\varphi} \textrm{f} $. The frequency $\nu$ is set indirectly by setting the delay time between two consecutive images $\Delta t_0$, that can be determined with a precision of $0.01\ ms$. This introduces a rounding error $\delta \nu_{round}/\textrm{f}\sim 10^{-5} $ of the same order of magnitude as the intrinsic shear frequency variability (see ESI for details). The parameter $m_\varphi$ has been chosen in order to minimize the rounding error. Since different shear amplitudes lead to a different actual frequency, the number of images per period may vary between experiments at different amplitude, taking values in the interval $m_{\varphi} \in [4,8]$.

\textbf{Residual drift and registration:}
The residual frequency mismatch results in a phase-dependent apparent drift (Fig.\ref{fgr:beat}) that is quite evident  in long acquisitions. By means of the same image cross-correlation algorithm used for the determination of the displacement field (see Sect. \ref{strainfieldmeas}), we characterize the residual drift, and we remove its effect with a custom sub-pixel registration algorithm implemented in the Fourier space and described in detail in {ESI}. In the case of fluid samples, actual drifts, corresponding to macroscopic flows, could also occur as a consequence of the loose lateral confinement. With the aforementioned registration algorithm, we can effectively remove these global contributions as well. 

\subsubsection{Summary of the echo strategy}\label{strategy}
In summary, our optimized strategy consists in acquiring a timed image sequence $I(\mathbf{x},t)$ including exactly $m_\varphi$ images per period. For each experiment at a given amplitude, we have $m_\varphi$ different stacks of echo images, that we label with their initial phase $\varphi_m=2\pi m/m_\varphi$, $m=0,1,...,(m_\varphi-1)$
\begin{equation}
\label{echoseq}
    	I_{n}(\mathbf{x}|\varphi_m)=I(\mathbf{x},(n+\varphi_m /2\pi)T).
\end{equation}
Registration and computation of the dynamic structure function are then sequentially performed on each stack separately by working at constant $m$ and calculating the image structure function for various delay cycles indexed by $n$.  Finally, the dynamic structure functions obtained for different initial phases are averaged to give a single dynamic structure function $D(\mathbf{q},\Delta t)$, representative of the average response of the sample within the cycle.
\subsubsection{Echo-Particle Tracking}
As a cross-check  of  the  results  obtained  with  echo-DDM,  we also perform a direct space analysis of the same image sequences, at least in those cases where the size of the tracer particles is large enough to enable them to be tracked individually.
For each acquisition, echo image sequences are extracted and pre-processed as described in Sect. \ref{strategy}. Images in each echo sequence are convoluted with a Wiener kernel (with standard deviation of $2 \mu$m) to reduce noise and then analyzed with a customized version of the MATLAB particle-tracking code developed by the Kilfoil group at the University of Massachusetts \cite{PhysRevLett.102.188303} and available at https://github.com/dsseara/microrheology. We extract the 
trajectories $[x^{(i)}(t),y^{(i)}(t)]$ of all particles (typically $\sim 100$) in the field of view, and we evaluate the corresponding displacement probability distribution functions (PDF) $P_s(\Delta x,\Delta t)$ and $P_v(\Delta y,\Delta t)$, along the shear ($x$) and the vorticity ($y$) direction, respectively
\begin{equation*}
P_s(\Delta x,\Delta t)=\langle\delta[\Delta x-(x^{(i)}(t+\Delta t)-x^{(i)}(t))]\rangle,
\end{equation*}
\begin{equation*}
P_v(\Delta y,\Delta t)=\langle\delta[\Delta y-(y^{(i)}(t+\Delta t)-y^{(i)}(t))]\rangle.
\end{equation*}
In the above expressions, the average is performed over all the initial times $t$, all the tracked particles $i$ and all the echo sequences corresponding to different initial phases. From the same data, we also calculate the mean square particle displacement (MSD) along  the shear and the vorticity direction, respectively
\begin{equation*}
MSD_s(\Delta t)=\langle|x^{(i)}(t+\Delta t)-x^{(i)}(t)|^2\rangle,
\end{equation*}
\begin{equation*}
MSD_v(\Delta t)=\langle|y^{(i)}(t+\Delta t)-y^{(i)}(t)|^2\rangle.
\end{equation*}

\section{Results}
\label{results}
\subsection{Measurement of the deformation profiles}
\label{resdef}
For each of the four samples, we perform a detailed characterization of the deformation profile which is produced when an oscillatory shear deformation of different amplitude is imposed. %We discuss the general features of the tests and the results referring to Fig.\ref{fgr:ZScan}.

\subsubsection{Carbopol}
\begin{figure}[h]
\centering
  \includegraphics[width=9 cm]{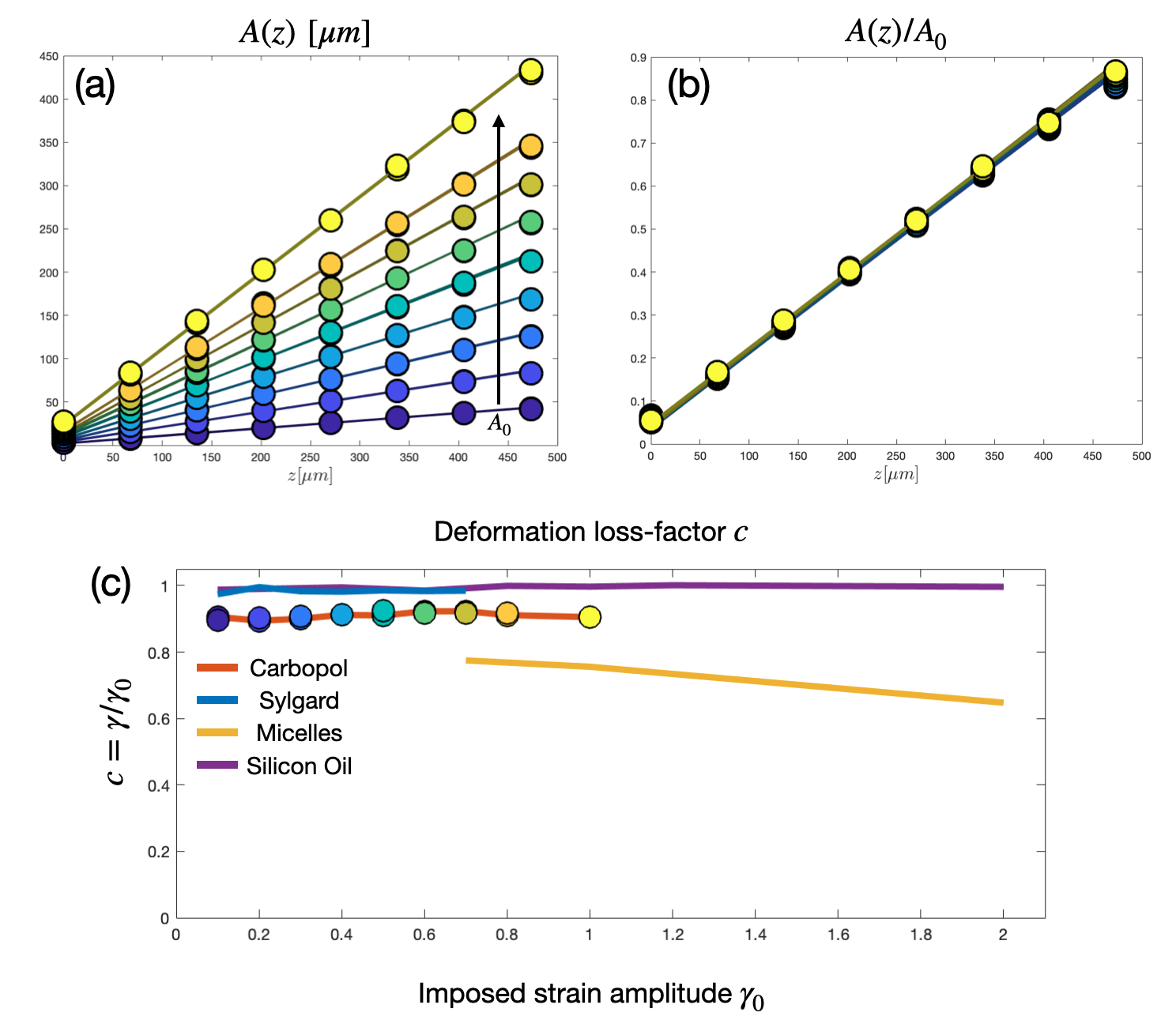}
  \caption{\textbf{Direct measurement of the deformation field}. (a) Displacement profile $A(z)$ in the Carbopol sample along the velocity gradient ($z$) direction for different applied strain amplitudes $\gamma_0 \in [10, 100] \%$ \textcolor{black}{(blue corresponds to $10\%$ and yellow to $100\%$)}. Symbols of different colors represent experimental data obtained at a different imposed shear amplitude, while lines are best fitting curves with the model of Eq.\ref{DFS}. For each amplitude, two independent, steady-state measurements are performed. In (b) the same profiles are rescaled with the amplitude $A_0$ of the imposed displacement. (c) {Deformation-loss-factor} for different samples  as a function of the imposed shear strain $\gamma_0$. Symbols correspond to Carbopol data. The color code is the same of panels (a-b).
 }
  \label{fgr:ZScan}
\end{figure}

In Fig.\ref{fgr:ZScan}(a), the results of vertical scans of the Carbopol sample for different shear amplitudes in the interval $\gamma_0\in[10, 100]\%$ are shown. For each amplitude (labeled by a different color) two successive full scans are performed. The excellent overlap between these pairs of measurements indicates that the observed deformation profiles do not significantly evolve over time and do not depend on the shear history. For all the imposed shear amplitudes, the displacement profile is almost perfectly linear. Fitting of each profile with Eq.\ref{DFS} enables estimating the corresponding deformation-loss-factor $c=\gamma/\gamma_0$, which quantifies the discrepancy between the imposed strain amplitude $\gamma_0$ and the measured one $\gamma$. In the case of Carbopol, the deformation-loss-factor is almost constant  $c\simeq 0.9$ (see Fig.\ref{fgr:ZScan}(c)), as can also be seen from the collapse of all the data onto a single master curve (panel (b)), which is obtained by normalizing each displacement profile by the displacement $A_0$ imposed on the top plate of the shear-cell.
The fact that the displacement profiles are linear for all the investigated strain amplitudes indicates that the material response remains homogeneous even in the transition region. This is a non-trivial result, as the occurrence of inhomogeneous weakening, accompanied by the formation of shear bands, is often observed for both hard materials and for other yield stress fluids when driven out of the linear regime  \cite{divoux2010, dinkgreve2018, joshi2018yield}.

\subsubsection{Sylgard} For this sample, representative of an ideal elastomer, with nearly  Hookean solid response, we performed three successive scans for each amplitude in the range $\gamma_0 \in [1\%, 70\%]$. Compared to Carbopol, the measured shear strain is in even better agreement with the imposed one ($c\simeq 0.98)$, as shown in Fig.\ref{fgr:ZScan}(c). The absence of wall slip is indicative of the strong adhesion between the gel sample and the confining plates, which was obtained by maintaining the sample in contact with the glass slides during the curing process. 
\subsubsection{Silicon oil} Very similar results are also obtained for the Newtonian sample. The reconstructed  deformation profile is homogeneous, and $c\simeq 0.98$ (see Fig.\ref{fgr:ZScan}(c)).
\subsubsection{Worm-like micelles}
Experiments performed with this sample reveal the presence of significant wall slip, which becomes increasingly stronger at large shear amplitudes, as indicated by the marked decrease of the deformation loss-factor as a function of the imposed deformation amplitude (yellow line in Fig.\ref{fgr:ZScan}(c)). Interestingly, for each amplitude, the reconstructed deformation profile is still homogeneous, characterized by a well-defined shear strain $\gamma<\gamma_0$ (see {ESI} for more details).

\subsection{Echo-DDM}
\label{resddm}
Once a detailed characterization of the sample deformation field is available, we proceed to quantify the shear-induced irreversible dynamics with the echo-DDM approach introduced in  Sect. \ref{EDDM}. As pointed out in Sect. \ref{technical}, a key requirement for the accurate determination of the non-affine, irreversible tracer displacements in the sample is the absence of spurious contributions (such as drifts or external vibrations) that could mask the genuine signal.  
Small imperfections or asymmetries in the setup may lead to a stress unbalance which, in the absence of an effective lateral confinement, can induce a net, albeit typically very small, sample flow. The relevance of this effect is found to crucially depend on the rheological properties of the material and it is particularly severe for low or no yield stress samples (see {ESI} for more details).
As shear-induced flows are often not perfectly homogeneous, they cannot be completely removed with the registration procedure described in Sect. \ref{technical}. As a matter of fact, for both the silicon oil and the WLM samples, the residual effect of shear-induced flows are found be large enough to dominate the microscopic dynamics (see {ESI} ), suggesting that our shear geometry is not optimal to measure irreversible tracer displacements in fluid-like samples. We stress that the effect under discussion has no impact on the results for the deformation profiles reported in Sect. \ref{resdef}, as the displacement per period associated with shear-induced flows is typically at least two orders of magnitude smaller than the amplitude of the affine displacement (see Fig.S5).

We thus focus on the two samples that display a solid-like behavior at rest: Sylgard and Carbopol. For these samples, all the echo acquisitions are performed at the center of the cell: $z=h/2$. This allows to neglect boundary effects (including the small wall slip) since the distance between the focal plane and the sample edges is way larger than the depth-of-focus $h/2\gg L_f$. \textcolor{black}{In principle, it would be also possible to perform measurements in the vicinity of the moving wall to study in more detail the plastic activity triggered by the interaction of the sample with the boundaries and, more generally, the role of surface boundary conditions for the flow of soft materials \cite{mansard2014boundary}}.
\subsubsection{Sylgard}
\begin{figure}[h]
\centering
  \includegraphics[width=9 cm]{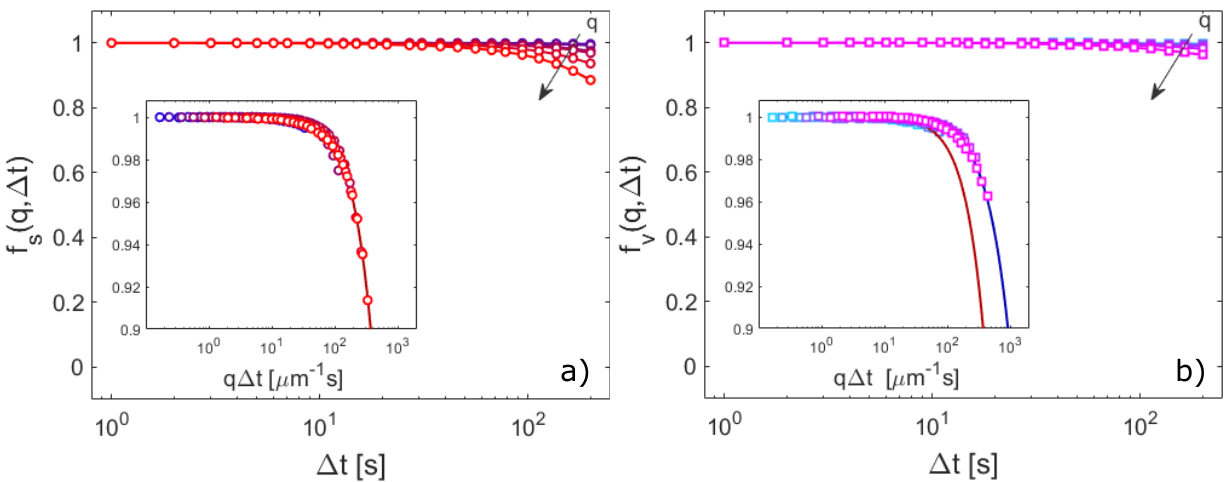}
  \caption
  {\textbf{Echo-DDM analysis of Sylgard. } 
 (a,b) Representative ISFs for different $q$-values in the range $[0.15-2.2]\mu m^{-1}$, along the shear and the vorticity directions, respectively. In the inset of panel (a), the same data shown in the main panel are plotted as a function of the rescaled time delay $q \Delta t$, displaying a nice collapse onto a single master curve (continuous red line). The same curve is also reported for comparison in the inset of panel (b), where the data shown in the corresponding panel are plotted as a function of $q \Delta t$, collapsing also in this case onto a single curve (continuous blue line). }
  \label{fgr:DDMSyl}
\end{figure}

This sample displays an elastic, reversible mechanical response over the whole explored range of shear amplitudes (up to $\gamma=80\%$) \cite{aime2016} and, as demonstrated above, it is firmly attached to the confining plates. Based on these premises, we do not expect to observe any genuine echo-DDM signal associated to irreversible non-affine displacements. As a consequence, these experiments can be thought of as a sort of calibration of the instrumental "background noise". For Sylgard, the ISFs obtained from an echo-DDM experiment with imposed shear amplitude $\gamma_0=75\%$ are shown in Fig.\ref{fgr:DDMSyl}.

As expected, the dynamics, along both the shear and the vorticity direction, is extremely slow and only the very first part of the decay of the ISFs can be observed. Nevertheless, when the ISFs are plotted as function of the rescaled delay time $q\Delta t$, a nice collapse of all the curves is observed (insets of Fig.\ref{fgr:DDMSyl}), suggesting the presence of a ballistic-like relaxation mechanism \cite{giavazzi2014}.
Assuming an exponential-like decay $f(q,\Delta t)\simeq e^{-vq\Delta t}$, we can get a rough, order-of-magnitude estimate of the characteristic associated velocity along the shear direction $v_s\sim 3\cdot 10^{-4}$ $\mu$m/s. 

We attribute this ultra-slow dynamics to the presence of an apparent relative motion of the different planes contributing to the image, which is due to the combination of imperfect synchronization and axial velocity gradient (see {ESI} for details). Consistently with this interpretation, we found that $v_b$ compares fairly well with the characteristic velocity spread $2\pi \delta \nu \gamma_0 L_f \sim 5 \cdot 10^{-4}$ expected in the presence of a synchronization error $\delta \nu \sim 10^{-5}$ Hz (see Sect.\ref{technical}) and a finite depth-of-focus $L_f\sim 10$ $\mu$m.
Moreover, we observe that the characteristic velocity related with the ballistic decay of the ISF along the vorticity direction is markedly slower (by about a factor of three, see inset of Fig.\ref{fgr:DDMSyl}(b)) compared to $v_s$. This is consistent with the presence in the echo images of an apparent ballistic motion mainly directed along the shear direction and marginally contributing to the dynamics in the perpendicular direction due to the finite angular aperture of the bow-tie regions used to calculate $f_v(q,\Delta t)$ (see Sect.\ref{DDM}).

These results show that the residual effects of imperfect  synchronization of the echo dynamics are rather small, corresponding to characteristic displacements of less than 1 nm per cycle. As it will be explicitly demonstrated in the next subsection, as soon as a genuine contribution to the dynamics, due to structural rearrangements, is present, the ballistic term discussed in this subsection can be safely neglected.

\subsubsection{Carbopol}
This sample is a yield stress fluid. As such, it exhibits a transition from solid to liquid, as the amplitude of the applied shear strain exceeds a certain threshold. This threshold can be broadly identified with the cross-over point of $G'$ and $G''$ in a dynamic strain sweep oscillatory rheological test, which we take here as an estimate of the sample yield strain \cite{dinkgreve2016} {(see  ESI for details)}. We thus expect the Carbopol sample to exhibit a dramatically different dynamics below and above the yield strain $\gamma_C\simeq 60\%$. \textcolor{black}{While a systematic study of this transition is surely interesting, to assess the potential of our methodology we selected two different shear amplitudes $\gamma$, respectively below ($\gamma_0=20\%$) and above ($\gamma_0=80\%$) the yield strain. In both cases, after the sudden application of the oscillatory shear, the sample exhibited a transient, whose effects were made negligible by waiting $120$ oscillation cycles before starting the image acquisition.}

\begin{figure*}[tbh!]
    \centering
  \includegraphics[width=.9 \linewidth]{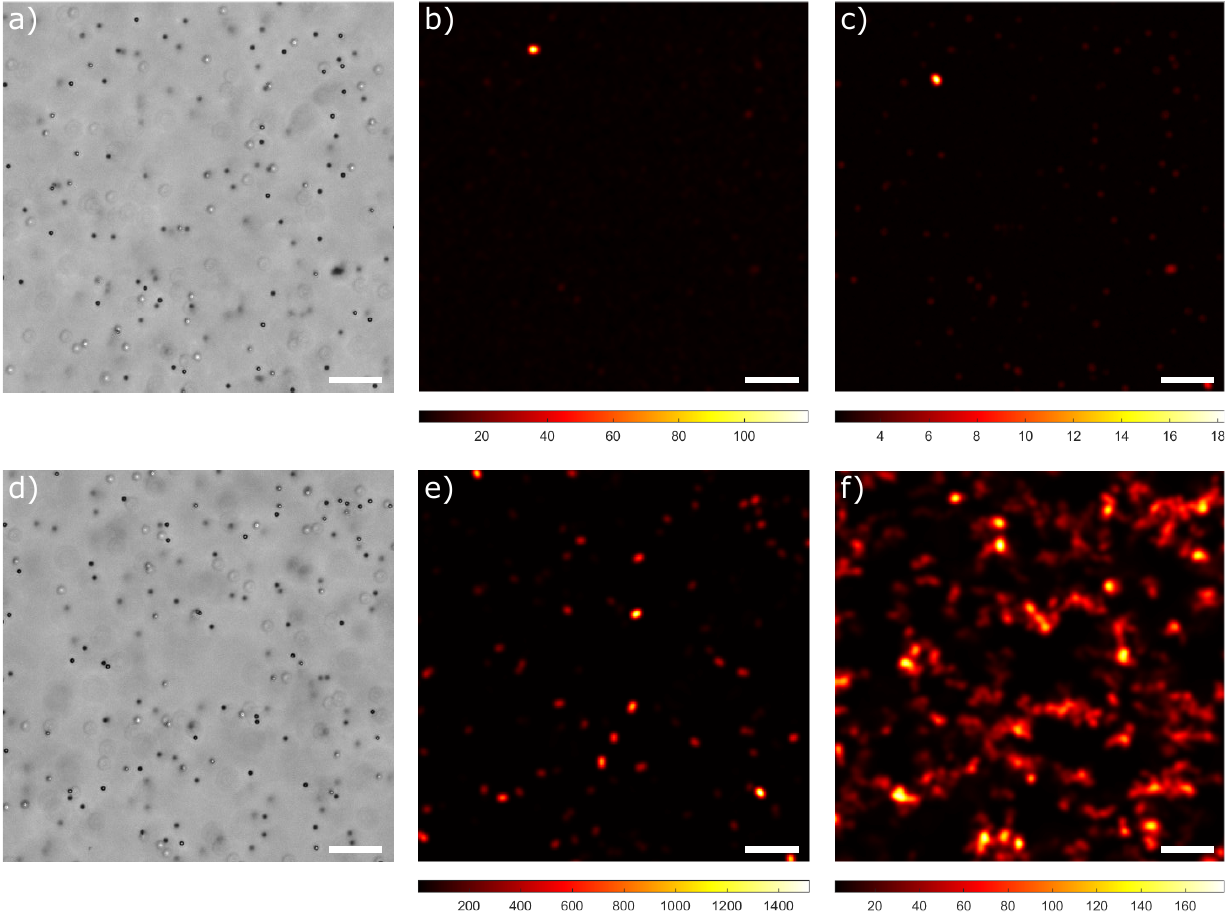}
  \caption{\textbf{Echo activity maps of the Carbopol sample.} {Panels (a-c) refer to an echo experiment with imposed shear strain $\gamma_0=20\%$, while panels (d-f) correspond to $\gamma_0=80\%$.   (a) Representative bright-field image of the tracers-seeded Carbopol sample. (b) Instantaneous activity map capturing one of the rare, localized rearrangement events occurring in the sheared sample for $\gamma_0=20\%$. (c) Integrated activity map outlining the presence of few, localized "active regions" corresponding to displacing tracers in an otherwise non-rearranging background. Panels (d-f) show the same quantities of panels (a-c) for $\gamma_0=80\%$. In all panels the scale bar corresponds to $35 \mu m$\textcolor{black}{, and the horizontal (vertical) axis represents the shear (vorticity) direction.}}}
  \label{fgr:Activity_lowhigh}
\end{figure*}

\textbf{Below the crossover point}\\
When the imposed amplitude is small ($\gamma_0=20\%$), visual inspection of the echo image sequences reveals that most of the tracers do not move appreciably, while a few of them undergo rare, intermittent displacements {(see supplementary movie M1)}. The occurrence of these isolated rearrangement events can be highlighted by considering instantaneous (Fig.\ref{fgr:Activity_lowhigh}b) or integrated echo activity maps (Fig.\ref{fgr:Activity_lowhigh}c). Instantaneous echo activity map are obtained as the square modulus of the difference between two consecutive images within an echo sequence $|I_{n}(\mathbf{x})-I_{n+1}(\mathbf{x})|^2$, while the integrated map corresponds to the time average of all the instantaneous maps in the sequence, and is a proxy of the cumulative rearrangements occurred in the sample during the whole experiment.

\begin{figure*}[tbh!]
    \centering
  \includegraphics[width=1 \linewidth]{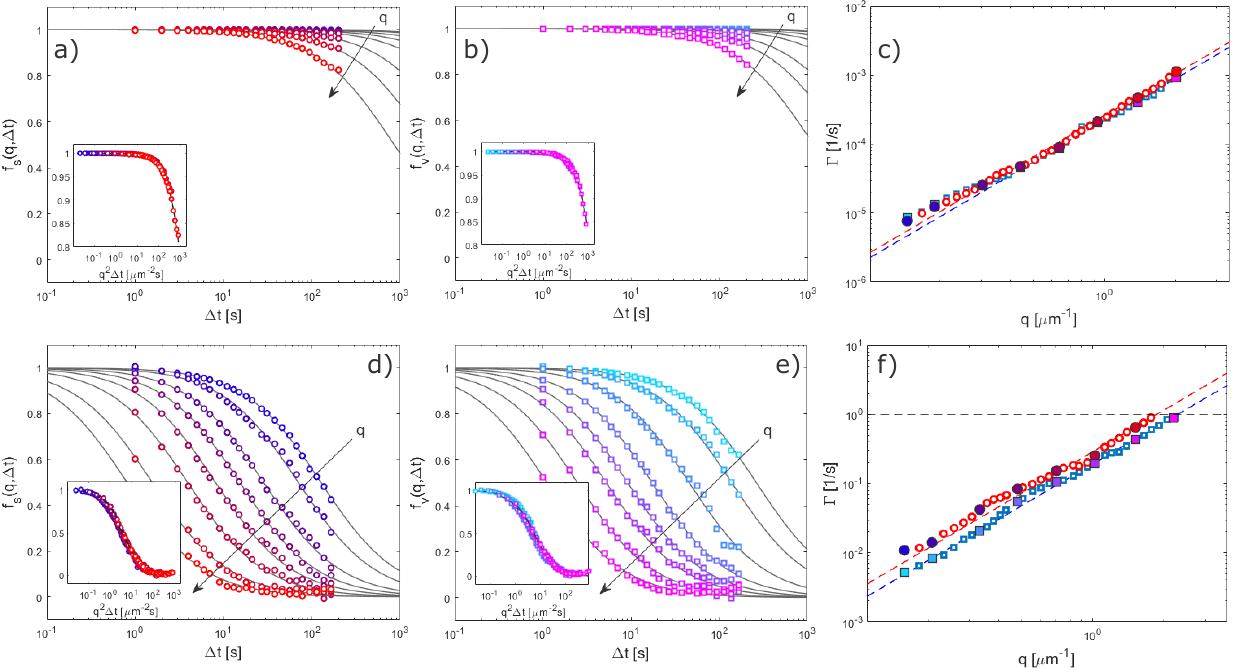}
 \caption{\textbf{Echo-DDM analysis of Carbopol. } Panels (a-c) refer to an echo experiment with imposed shear strain $\gamma_0=20\%$, while panels (d-f) correspond to $\gamma_0=80\%$.
 {(a,b) Representative ISFs for different $q$-values in the range $[0.15-2.2]\mu m^{-1}$, along the shear and the vorticity directions, respectively.
Continuous lines are best fitting curves to the experimental data with the model $f(q,\Delta t) = [1+\Delta t\Gamma(q)]^{-1}$, from which an estimate of the $q$-dependent relaxation rate $\Gamma(q)$ is obtained. In the inset of each panel, the same data of the main figure are plotted as a function of the rescaled time delay $q^2 \Delta t$. (c) $q$-dependent relaxation rates $\Gamma_s(q)$ (orange circles) and $\Gamma_v(q)$ (blue circles), along the shear and vorticity directions, respectively. Large circles correspond to $q$-values whose ISFs are shown in panels (a-b), with the same colors. Dashed lines represent the best fits to the relaxation rates with a quadratic model $\Gamma(q)=Dq^2$. (d-f) same of (a-c) for imposed shear amplitude $\gamma_0=80\%$. In (f) the horizontal black dashed line marks the maximum relaxation rate that can be reliably measured, corresponding to the oscillation frequency $\nu =1Hz$ .}}
  \label{fgr:DDM_lowhigh}
\end{figure*}

To be more quantitative, we apply echo-DDM by considering separately the dynamics along the shear and the vorticity direction. As it can be seen from Fig.\ref{fgr:DDM_lowhigh}(a,b), for this amplitude we cannot capture the full relaxation of the ISFs within our observation window. Nevertheless, as shown in the inset of panels (a,b), all the ISFs obtained for $q\in[0.15-2.2]$ $\mu m^{-1}$ nicely collapse onto a single master curve when plotted as a function of the rescaled time delay $q^2 \Delta t$. This indicates that even below yielding the system undergoes an (extremely slow) diffusive-like relaxation, which results from irreversible rearrangements. This is in agreement with the seminal observations made with DWS in Ref. \cite{hebraud1997}. Here, we have the advantage that we can characterize this irreversible dynamics as a function of the wave-vector $q$ on a well-defined plane in the sample, while being able to simultaneously measure the local strain. In order to measure the $q$-dependent relaxation rates, we thus fitted the phenomenological model $f(q,\Delta t) = [1+\Delta t\Gamma(q)]^{-1}$, to the experimental ISFs. According to Eq.\ref{ISF_MSD}, this model provides the Fickean scaling  $\langle\Delta x^2(\Delta t)\rangle = 2D_0\Delta t$ for the 1D MSD when the DDM relaxation rates exhibit a quadratic dependence $\Gamma(q)= D_0 q^2$ on the wavevector. This is exactly the scaling that we observe in both directions, with a moderate deviation only for the smallest values of $q$ (Fig.\ref{fgr:DDM_lowhigh}(c)). By fitting the relaxation rates we obtain the estimates $D_s=(2.2 \pm 0.2)\cdot 10^{-4} \mu m^2/s$ and $D_v=(2.6 \pm 0.2)\cdot 10^{-4} \mu m^2/s$ for the diffusion coefficient associated with the shear-induced effective dynamics of the tracers along the shear and the vorticity direction, respectively. Within the experimental error, the shear-induced dynamics is isotropic $D_s\simeq D_v$ with values for the diffusion coefficients about one order of magnitude larger than the one associated with the thermal motion of the same tracers in Carbopol under stationary conditions (see ESI for details). \textcolor{black}{Beyond being slower, the tracer dynamics in the sample at rest is also non-intermittent, which contrasts with the rare, intermittent rearrangements that can be spotted in the sample sheared below the yield strain. Interestingly, this small number of intermittently rearranging particles coexist with a larger number of slowly, yet persistently rearranging ones, with the result that the overall tracer dynamics remains diffusive. In these conditions, the determination of the diffusion coefficient is sensitive to the choice of the field of view, and it would be interesting to study the spatial correlation between rearrangements. A detailed study of these effects goes beyond the scope of the present study}.

%We note that the adopted functional form of the dispersion relation is compatible with the interpretation of the parameter $D$ as a single particle diffusion coefficient. In fact, according to Eq.\ref{ISF_MSD}, this choice provides the expected scaling  $\langle\Delta x^2(\Delta t)\rangle = 2D\Delta t$ for the 1D MSD. 

\begin{figure*}[tbh!]
    \centering
  \includegraphics[width=1 \linewidth]{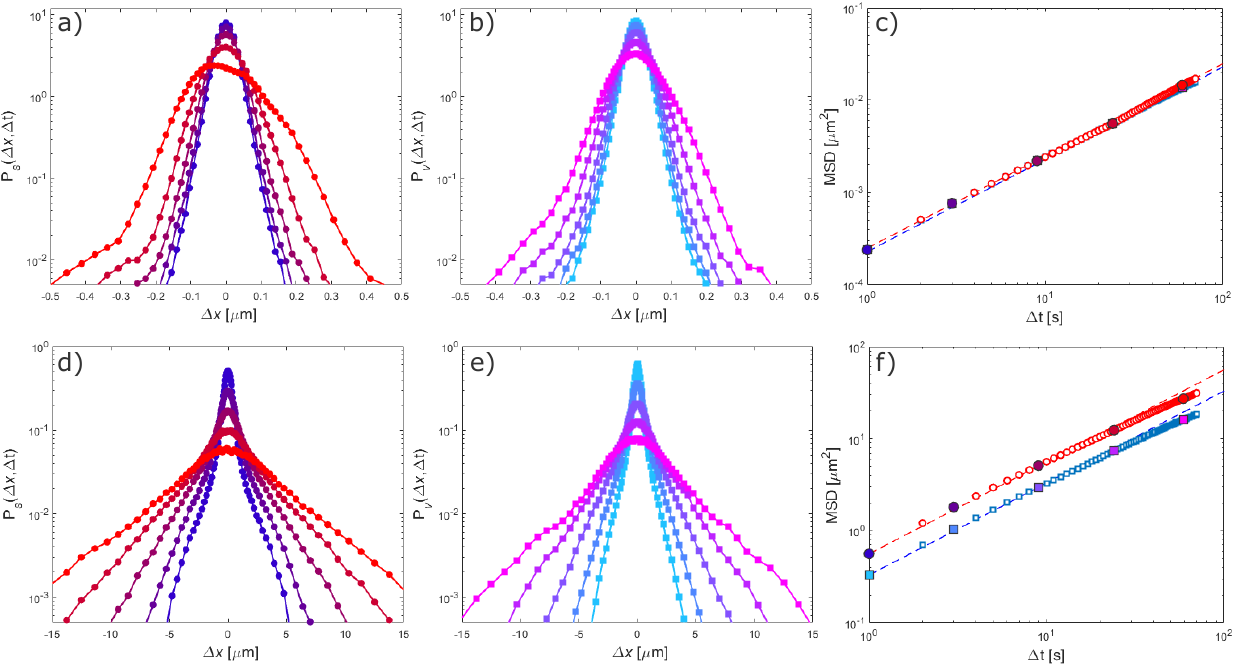}
  \caption
  {\textbf{Echo-PT analysis of Carbopol.} Panels (a-c) refer to an echo experiment with imposed shear strain $\gamma_0=20\%$, while panels (d-f) correspond to $\gamma_0=80\%$. (a,b) Probability density functions (PDFs) of particle displacements $P_s(\Delta \mathbf{x},\Delta t)$ and $P_v(\Delta \mathbf{x},\Delta t)$ for different time delays $\Delta t$ in the range $[1, 70]$ s evaluated along the shear and the vorticity direction, respectively. (c) Symbols: mean square displacements of the particles along the shear (red) and the vorticity (blue) direction, respectively.  Large circles correspond to the delay times whose PDFs are shown in panels (a-b), with the same colors.
  The red (blue) dashed line is the best fitting curve to the MSD along the shear (vorticity) direction with a linear model $MSD(\Delta t)=2D\Delta t$. (d-f) same of (a-c) for imposed shear amplitude $\gamma_0=80\%$.}
  \label{fgr:MSD_lowhigh}
\end{figure*}

The results obtained with DDM on the echo image sequences are confirmed by PT analysis of the same sequences. The estimated PDF of the tracer displacements along the shear and the vorticity direction, is shown in Fig.\ref{fgr:MSD_lowhigh}(a) and (b), respectively. In agreement with DDM analysis, we do not find any statistically significant difference between the shear and the vorticity direction, confirming that the dynamics is substantially isotropic. Interestingly, for all investigated time delays, the PDFs is not Gaussian, as one would expect for ordinary diffusion, but displays exponential-like tails.
Nevertheless, the MSDs in the shear and vorticity directions, reported in panel (c), display a clean linear dependence on the delay time,  $MSD_s(\Delta t)\simeq 2 D_{PT,v} \Delta t$, $MSD_v (\Delta t)\simeq 2 D_{PT,v} \Delta t$, indicating Fickean diffusion. The value of the associated diffusion coefficient is found to be $D_{PT,s}=(1.14 \pm 0.08)\cdot 10^{-4} \mu m^2/s$ and $D_{PT,v}=(1.24 \pm 0.1)\cdot 10^{-4} \mu m^2/s$ respectively. The discrepancy (about a factor of 2) between the estimates for the diffusion coefficient obtained from DDM and PT can be probably attributed to differences in the populations whose dynamics is captured by the two techniques (DDM sees all the particles and PT only a user-dependent selection).

\textbf{Above the crossover point}\\
When the sample is subjected to a large amplitude shear deformation $\gamma_0=80\%$, exceeding the crossover threshold $\gamma_C$, a dramatic change in the echo dynamics is observed. The amplitude of tracer displacements increases by orders of magnitude and the shear-induced activity shows a much larger degree of spatial homogeneity and temporal uniformity {(see supplementary movie M2)}. 
This can be also appreciated by inspecting the echo activity maps reported in Fig.\ref{fgr:Activity_lowhigh}(e-f), outlining the presence of plastic, irreversible rearrangements distributed across the whole field of view. 
The speed-up of the dynamics is well captured by echo-DDM analysis. In contrast with the small deformation case, the obtained ISFs display an almost complete temporal relaxation for all the investigated $q$-values, in both the shear and the vorticity direction (see Fig.\ref{fgr:DDM_lowhigh}(d-e)). By plotting the ISFs as a function of the rescaled time delay $q^2\Delta t$, an excellent collapse of the ISFs on a single master curve is observed, highlighting once again a diffusive-like dynamics (inset of panels (d-e)).
As it can be appreciated from Fig.\ref{fgr:DDM_lowhigh}(d-e), over the whole considered $q$-range, the decay of ISFs is very well captured by the same one-parameter fitting model $f(q,\Delta t) = [1+\Delta t\Gamma(q)]^{-{1}}$ used for the small deformation case. The obtained $q$-dependent relaxation rates show a quadratic scaling with $q$ along both directions (Fig.\ref{fgr:DDM_lowhigh}f). Remarkably, in this case the dynamics is no longer isotropic, displaying a faster relaxation along the shear direction. Indeed, fitting the experimental relaxation rates to $\Gamma_v(q)=D_v q^2$ and $\Gamma_s(q)=D_sq^2$ provides $D_v=(1.9 \pm 0.3)\cdot 10^{-1} \mu m^2s^{-1}$ and $D_s = (2.9 \pm 0.3)\cdot 10^{-1} \mu m^2s^{-1}$, respectively. Comparing the diffusion coefficients for $\gamma_0=20\%$ and $\gamma_0=80\%$, we observe a $\sim1000$-fold speed-up of the dynamics for large amplitude deformation, which is a strong signature of shear-induced diffusion.
PT-based analysis of the same data fully corroborates the results obtained with echo-DDM, and in this case the agreement with DDM is even better than for small deformation. In particular, they exhibit a linear dependence on $\Delta t$. From a linear fit $MSD=2D\Delta t$ of the MSDs along both the investigated directions (Fig.\ref{fgr:MSD_lowhigh}f) we obtain $D_{PT,v}=(1.7\pm 0.2)\cdot 10^{-1} \mu  m^2s^{-1}$ and $D_{PT,s}=(2.8\pm 0.2)\cdot 10^{-1} \mu  m^2s^{-1}$, respectively, in fairly good agreement with the results obtained from echo-DDM.

\section{Conclusion}
\label{conklu}
In this paper, we combined oscillatory deformation in a simple shear geometry with bright-field optical microscopy.
For four different samples, we characterized the deformation field, finding stationary and homogeneous deformation fields. We observed the presence of partial slip, and we measured the actual local deformation.
In echo measurements,  we observed drifts of different origin: "apparent" drifts due to an imperfect synchronization between the shear and the acquisition frequencies, and "real" drifts due to unavoidable instrumental imperfections and the lack of horizontal confinement. The effect of drift on the microscopic dynamics is corrected with a custom sub-pixel registration algorithm.
Eventually, the DDM algorithm is applied to echo movies.
In the case of an "ideal" elastic solid, the applied procedure shows, as expected, the absence of shear-induced-dynamics. In the fluid-like samples, the dynamics is dominated by flows generated by small asymmetry in the set-up and by the lack of lateral confinement of the samples. Although our registration algorithm can significantly reduce the impact of these drifts, their residual effect is still large enough to prevent the observation of any additional shear-induced dynamics.

With Carbopol, we perform experiments by using two different strain amplitudes, below and above the yield strain. The smaller amplitude causes a spatially heterogeneous, temporally intermittent activity of the tracer particles, which results in extremely slow, substantially isotropic shear-induced diffusion. Above the yield strain, the shear-induced diffusive relaxation becomes anisotropic and very fast, with an observed speed-up of about $10^3$ compared to the previous case. Our results with Carbopol represent the first clear observation of a truly diffusive dispersion relation $\Gamma(q)\sim q^2$ in fluids that are yielding during oscillatory experiments. Of note, this behavior contrasts with recent results obtained with x-ray photon correlation spectroscopy \cite{rogers2014, rogers2018microscopic} in which a faster-than-diffusive relaxation (closer to the behavior observed in colloidal polycrystals \cite{tamborini2014}) was found for a nanoemulsion and a colloidal gel; at the same time, our results agree with the original interpretation provided for dense emulsions \cite{hebraud1997} and colloidal glasses \cite{petekidis2002,petekidis2003}, in both cases lacking an experimental determination of the dispersion relation.

Beyond providing reliable and quantitative information about the shear-induced microscopic rearrangements, our approach also enables characterizing the deformation profiles across the sample and, in turn, spotting slippage, shear bands or other unwanted effects. This is made possible by the partial coherence of the illuminating light, which not only helps in identifying the presence of the above-mentioned phenomena, but also allows imaging a selected slice of the sample. In this work, we imaged its central part, where the deformation is homogeneous, by excluding the sample boundaries where deviations from homogeneity (mainly due to wall slip) possibly occur. In case of shear-banding, the $z$-selectivity of the microscope can be exploited to select zones where the macroscopic deformation field is approximately homogeneous, an important advantage over traditional single and multiple light scattering approaches. \textcolor{black}{For this reason, our technique appears promising for the study of the plastic dynamics in the presence of shear banding, the interplay between non-affine displacements and shear banding representing a very interesting, yet not much explored topic.}. Another advantage of our approach is that it makes it possible to combine real-space pre-processing (drift correction) and Fourier-space analysis (Echo-DDM), which leads to a considerable simplification of the data interpretation.

Extensions of our approach to other imaging modes (\textit{e.g.} confocal fluorescence microscopy) are easily within reach, as well as combination with a commercial rheometer provided that the sample remains optically accessible. Finally, the usual advantages (and disadvantages) of DDM over PT hold, including the capability of DDM to probe the dynamics of sub-resolution particles \cite{edera2017}.

\section*{Conflicts of interest}
There are no conflicts to declare.
\section*{Acknowledgements}
We acknowledge useful discussions with Manuel Escobedo, Stefan Egelhaaf, and Veronique Trappe. \textcolor{black}{The work benefited from support from the European Soft Matter Infrastructure (EUSMI).}

%%%END OF MAIN TEXT%%%
\balance
%%%REFERENCES%%%
\bibliography{rsc.bib} %You need to replace "rsc" on this line with the name of your .bib file
\bibliographystyle{rsc} %the RSC's .bst file

\end{document}